\documentclass[twocolumn,superscriptaddress,preprintnumbers,amsmath,amssymb]{revtex4}
\usepackage{graphicx}% Include figure files
\usepackage{dcolumn}% Align table columns on decimal point
\usepackage{bm}% bold math

\begin{document}

%\preprint{APS/123-QED}

\title{ Pion Interferometry in Au+Au and Cu+Cu Collisions at 
$\sqrt{s_{\rm{NN}}}$ = 62.4 and 200 GeV}

\affiliation{Argonne National Laboratory, Argonne, Illinois 60439, USA}
\affiliation{University of Birmingham, Birmingham, United Kingdom}
\affiliation{Brookhaven National Laboratory, Upton, New York 11973, USA}
\affiliation{University of California, Berkeley, California 94720, USA}
\affiliation{University of California, Davis, California 95616, USA}
\affiliation{University of California, Los Angeles, California 90095, USA}
\affiliation{Universidade Estadual de Campinas, Sao Paulo, Brazil}
\affiliation{University of Illinois at Chicago, Chicago, Illinois 60607, USA}
\affiliation{Creighton University, Omaha, Nebraska 68178, USA}
\affiliation{Nuclear Physics Institute AS CR, 250 68 \v{R}e\v{z}/Prague, Czech Republic}
\affiliation{Laboratory for High Energy (JINR), Dubna, Russia}
\affiliation{Particle Physics Laboratory (JINR), Dubna, Russia}
\affiliation{Institute of Physics, Bhubaneswar 751005, India}
\affiliation{Indian Institute of Technology, Mumbai, India}
\affiliation{Indiana University, Bloomington, Indiana 47408, USA}
\affiliation{Institut de Recherches Subatomiques, Strasbourg, France}
\affiliation{University of Jammu, Jammu 180001, India}
\affiliation{Kent State University, Kent, Ohio 44242, USA}
\affiliation{University of Kentucky, Lexington, Kentucky, 40506-0055, USA}
\affiliation{Institute of Modern Physics, Lanzhou, China}
\affiliation{Lawrence Berkeley National Laboratory, Berkeley, California 94720, USA}
\affiliation{Massachusetts Institute of Technology, Cambridge, MA 02139-4307, USA}
\affiliation{Max-Planck-Institut f\"ur Physik, Munich, Germany}
\affiliation{Michigan State University, East Lansing, Michigan 48824, USA}
\affiliation{Moscow Engineering Physics Institute, Moscow Russia}
\affiliation{City College of New York, New York City, New York 10031, USA}
\affiliation{NIKHEF and Utrecht University, Amsterdam, The Netherlands}
\affiliation{Ohio State University, Columbus, Ohio 43210, USA}
\affiliation{Old Dominion University, Norfolk, VA, 23529, USA}
\affiliation{Panjab University, Chandigarh 160014, India}
\affiliation{Pennsylvania State University, University Park, Pennsylvania 16802, USA}
\affiliation{Institute of High Energy Physics, Protvino, Russia}
\affiliation{Purdue University, West Lafayette, Indiana 47907, USA}
\affiliation{Pusan National University, Pusan, Republic of Korea}
\affiliation{University of Rajasthan, Jaipur 302004, India}
\affiliation{Rice University, Houston, Texas 77251, USA}
\affiliation{Universidade de Sao Paulo, Sao Paulo, Brazil}
\affiliation{University of Science \& Technology of China, Hefei 230026, China}
\affiliation{Shandong University, Jinan, Shandong 250100, China}
\affiliation{Shanghai Institute of Applied Physics, Shanghai 201800, China}
\affiliation{SUBATECH, Nantes, France}
\affiliation{Texas A\&M University, College Station, Texas 77843, USA}
\affiliation{University of Texas, Austin, Texas 78712, USA}
\affiliation{Tsinghua University, Beijing 100084, China}
\affiliation{United States Naval Academy, Annapolis, MD 21402, USA}
\affiliation{Valparaiso University, Valparaiso, Indiana 46383, USA}
\affiliation{Variable Energy Cyclotron Centre, Kolkata 700064, India}
\affiliation{Warsaw University of Technology, Warsaw, Poland}
\affiliation{University of Washington, Seattle, Washington 98195, USA}
\affiliation{Wayne State University, Detroit, Michigan 48201, USA}
\affiliation{Institute of Particle Physics, CCNU (HZNU), Wuhan 430079, China}
\affiliation{Yale University, New Haven, Connecticut 06520, USA}
\affiliation{University of Zagreb, Zagreb, HR-10002, Croatia}

\author{B.~I.~Abelev}\affiliation{University of Illinois at Chicago, Chicago, Illinois 60607, USA}
\author{M.~M.~Aggarwal}\affiliation{Panjab University, Chandigarh 160014, India}
\author{Z.~Ahammed}\affiliation{Variable Energy Cyclotron Centre, Kolkata 700064, India}
\author{B.~D.~Anderson}\affiliation{Kent State University, Kent, Ohio 44242, USA}
\author{D.~Arkhipkin}\affiliation{Particle Physics Laboratory (JINR), Dubna, Russia}
\author{G.~S.~Averichev}\affiliation{Laboratory for High Energy (JINR), Dubna, Russia}
\author{J.~Balewski}\affiliation{Massachusetts Institute of Technology, Cambridge, MA 02139-4307, USA}
\author{O.~Barannikova}\affiliation{University of Illinois at Chicago, Chicago, Illinois 60607, USA}
\author{L.~S.~Barnby}\affiliation{University of Birmingham, Birmingham, United Kingdom}
\author{J.~Baudot}\affiliation{Institut de Recherches Subatomiques, Strasbourg, France}
% ...

\collaboration{STAR Collaboration}\noaffiliation

%\author{STAR Collaboration}

\date{\today}% It is always \today, today,

\begin{abstract}
We present a systematic analysis of two-pion interferometry in Au+Au collisions 
at $\sqrt{s_{\rm{NN}}}$ = 62.4 GeV and Cu+Cu collisions at $\sqrt{s_{\rm{NN}}}$ = 62.4 
and 200 GeV using the STAR detector at RHIC. The multiplicity and transverse momentum 
dependences of the extracted correlation lengths (radii) are studied. The scaling with charged particle multiplicity of the apparent 
system volume at final interaction is studied for the RHIC energy 
domain. The multiplicity scaling of the measured correlation radii is found to be 
independent of colliding system and collision energy. 
\end{abstract}

\maketitle

\section{Introduction}
\label{intro}

One of the definitive predictions of quantum chromodynamics (QCD) 
is that at sufficiently high temperature or density~\cite{McLerran:1980pk}
strongly interacting matter will be in a state with colored degrees 
of freedom, i.e. quarks and gluons. The central goal of the experiments 
with relativistic heavy ion collisions is to create and study this  
hypothesized form of matter, called the quark-gluon plasma (QGP),
which might have existed in the microsecond old universe. 
Numerous experimental observables have been proposed as signatures of 
QGP creation in heavy ion collisions~\cite{Adams:2005dq}. 
One of these predictions is based on the expectation that the increased number of degrees 
of freedom associated with the color deconfined 
state increases the entropy of the system which 
should survive subsequent hadronization and freeze-out (final interactions). 
The increased entropy is expected to lead to an increased spatial extent and duration of particle emission,
thus providing a significant probe for the QGP phase transition~\cite{Rischke:1996em,Rischke:1996nq}.

The information about the space-time structure of the emitting source 
can be extracted with intensity interferometry 
techniques~\cite{Goldhaber:1960sf}. This method, popularly known as
Hanbury Brown and Twiss (HBT) correlations, was originally developed to measure 
angular sizes of stars~\cite{HanburyBrown:1954wr}. The momentum correlations 
of the produced particles from hadronic sources however include dynamical as
well as interference effects, hence the term {\it femtoscopy}~\cite{Lednicky:2002fq} 
is more appropriate. The primary goal of femtoscopy, performed at 
mid-rapidity and low transverse momentum, is to study the space-time 
size of the emitting source and freeze-out processes of the dynamically 
evolving collision system. Femtoscopic correlations have been successfully studied 
in most of the heavy ion experiments (see ~\cite{Lisa:2005dd} for a recent review).

Experimentally, the two-particle correlation function is the ratio,
\begin{equation}
\label{eq:one}
C( \vec{q},\vec{K}) = 
\frac{A( \vec{q},\vec{K})}{B(\vec{q},\vec{K})}~~,
\end{equation}
where $A(\vec{q},\vec{K}$) 
is the distribution of pairs of particles with 
relative momentum $\vec{q} = \vec{p_1} - \vec{p_2}$ 
and average momentum $\vec{K} = (\vec{p_1} + \vec{p_2})/2$ 
from the same event, and $B(\vec{q},\vec{K}$) 
is the corresponding distribution for pairs of particles taken from 
different events~\cite{Kopylov:1972qw,Heinz:1999rw}. The correlation function
is normalized to unity at large $\vec{q}$. 
With the availability of high statistics data and development of new techniques,
it has become possible to measure three-dimensional
decompositions of $\vec{q}$~\cite{Bertsch:1988db,Pratt:1986cc,Chapman:1994yv},
providing better insight into the collision geometry.

Previous femtoscopic measurements at RHIC in Au+Au collisions at  
$\sqrt{s_{\rm{NN}}}$ = 130~GeV~\cite{Adler:2001zd,Adcox:2002uc} and 
200~GeV~\cite{Adler:2004rq,Adams:2004yc} obtained qualitatively similar source sizes. 
However, detailed comparisons with smaller colliding systems 
and energies are required in order to understand the dynamics of the source during freeze-out.
The crucial information provided from such femtoscopic studies with pions will help to 
improve our understanding of the reaction mechanisms and to constrain theoretical models of 
heavy ion collisions~\cite{Baym:1997ce,Padula:2004ba,Shuryak:1972kq,McLerran:1988rn,Heinz:1996bs,
Wiedemann:1999qn,Heinz:1996rw,Tomasik:2002rx}.

In this paper we present a systematic analysis of two-pion interferometry in Au+Au 
collisions at $\sqrt{s_{\rm{NN}}}$ = 62.4~GeV and Cu+Cu collisions at 
$\sqrt{s_{\rm{NN}}}$ = 62.4~GeV and 200~GeV using 
the Solenoidal Tracker at RHIC (STAR) detector at the 
Relativistic Heavy Ion Collider (RHIC). 
The article is organized as follows : Section~\ref{exptsetup} explains the 
detector set-up, along with the necessary event, particle and pair cuts. 
In Section~\ref{analysis}, the analysis and construction of the correlation function is 
discussed.
The presented results are compared with previous STAR measurements for Au+Au
collisions at $\sqrt{s_{\rm{NN}}}$ = 200~GeV in Section~\ref{results}. This section also includes 
a compilation of freeze-out volume estimates for all available 
heavy ion results from AGS, SPS and RHIC. 
Section~\ref{summary} contains a summary and conclusions.

\section{Experimental Setup, Event and Particle Selection}
\label{exptsetup}

\subsection{The STAR detector and Trigger details}
\label{stardet}

   The STAR detector~\cite{Ackermann:2002ad}, which has a large acceptance 
and is azimuthally symmetric, consists of several detector sub-systems and a solenoidal magnet.
In the present study the central Time Projection Chamber (TPC)~\cite{Ackermann:1999kc} provided the main information used for track 
reconstruction. It is 4.2 m long and 4 m in diameter. The TPC covers the pseudo-rapidity region $|{\eta}|$ $<$ 1.8 with full 
azimuthal coverage (-$\pi$ $<$${\phi}$$<$$\pi$ ). It is a gas chamber filled with P10 gas 
(10$\%$ methane, 90$\%$ argon) with inner and outer radii of 50 and 200 cm, respectively, 
in a uniform electric field of $\sim$ 135 V/cm. The paths of the particles passing through 
the gas are reconstructed from the release of secondary electrons that drift to the readout 
end caps at both ends of the chamber. The readout system is based on multi-wire proportional 
chambers with cathode pads. There are 45 pad-rows between the inner and outer radii of the TPC.

A minimum bias trigger is obtained using the charged particle hits from an array of 
scintillator slats arranged in a barrel, called the Central Trigger Barrel, surrounding 
the TPC, two Zero-Degree Calorimeters (ZDCs)~\cite{Adler:2000bd} at $\pm$18 m from the detector center along 
the beam line, and two Beam-Beam Counters. The ZDCs measure neutrons 
at beam rapidity which originate from the break-up of the colliding nuclei. The centrality determination 
which is used in this analysis is the uncorrected multiplicity of charged particles in the pseudo-rapidity 
region $|{\eta}|$ $<$ 0.5 ($N_{\rm ch}^{\rm TPC}$) as measured by the TPC.

\begin{table}
\caption{\label{tab:table1}Collision centrality selection in terms of percentage of total Au-Au inelastic cross-section, 
number tracks in TPC, average number of participating nucleons and average number of binary nucleon-nucleon collisions 
for Au+Au at $\sqrt{s_{\rm{NN}}}$ = 62.4 GeV.}
\begin{ruledtabular}
\begin{tabular}{cccc}
$\%$ cross-section&$N_{\rm ch}^{\rm TPC}$&$\langle N_{\rm part}\rangle$&$\langle N_{\rm coll} \rangle$\\
\hline\\
0-5   & $>$373  & $347.3_{-3.7}^{+4.3}$  & $904_{-62.4}^{+67.7}$\\\\
5-10  & 372-313 & $293.3_{-5.6}^{+7.3}$  & $713.7_{-54.8}^{+63.7}$\\\\
10-20 & 312-222 & $229.0_{-7.7}^{+9.2}$  & $511.8_{-48.2}^{+54.9}$\\\\
20-30 & 221-154 & $162.0_{-9.5}^{+10.0}$ & $320.9_{-39.2}^{+43.0}$\\\\
30-40 & 153-102 & $112.0_{-9.1}^{+9.6}$   & $193.5_{-31.4}^{+30.4}$\\\\
40-50 & 101-65  & $74.2_{-8.5}^{+9.0}$   & $109.3_{-21.8}^{+22.1}$\\\\
50-60 & 64-38   & $45.8_{-7.1}^{+7.0}$   & $56.6_{-14.3}^{+15.0}$\\\\
60-70 & 37-20   & $25.9_{-5.6}^{+5.6}$   & $26.8_{-9.0}^{+8.8}$\\\\
70-80 & 19-9    & $13.0_{-4.6}^{+3.4}$   & $11.2_{-4.8}^{+3.7}$\\\\
\end{tabular}
\end{ruledtabular}
\end{table}

\begin{table}
\caption{\label{tab:table2}
Collision centrality selection in terms of percentage of total Cu-Cu inelastic cross-section, 
number tracks in TPC, average number of participating nucleons and average number of binary nucleon-nucleon collisions
for Cu+Cu at $\sqrt{s_{\rm{NN}}}$ = 200 GeV.}
\begin{ruledtabular}
\begin{tabular}{cccc}
$\%$ cross-section&$N_{\rm ch}^{\rm TPC}$&$\langle N_{\rm part} \rangle$&$\langle N_{\rm coll} \rangle$\\
\hline\\
0-10   & $>$139  & $99.0_{-1.2}^{+1.5}$  & $188.8_{-13.4}^{+15.4}$\\\\
10-20  & 138-98  & $74.6_{-1.0}^{+1.3}$  & $123.6_{-8.3}^{+9.4}$\\\\
20-30  & 97-67   & $53.7_{-0.7}^{+0.9}$  & $77.6_{-4.7}^{+5.4}$\\\\
30-40  & 66-46   & $37.8_{-0.5}^{+0.7}$  & $47.7_{-2.7}^{+2.8}$\\\\
40-50  & 45-30   & $26.2_{-0.4}^{+0.5}$  & $29.2_{-1.4}^{+1.6}$\\\\
50-60  & 29-19   & $17.2_{-0.2}^{+0.4}$  & $16.8_{-0.7}^{+0.7}$\\\\
\end{tabular}
\end{ruledtabular}
\end{table}

\begin{table}
\caption{\label{tab:table3}Collision centrality selection in terms of percentage of total Cu-Cu inelastic cross-section, 
number tracks in TPC, average number of participating nucleons and average number of binary nucleon-nucleon collisions 
for Cu+Cu at $\sqrt{s_{\rm{NN}}}$ = 62.4 GeV.}
\begin{ruledtabular}
\begin{tabular}{cccc}
$\%$ cross-section&$N_{\rm ch}^{\rm TPC}$&$\langle N_{\rm part} \rangle$&$\langle N_{\rm coll} \rangle$\\
\hline\\
0-10   & $>$101  & $96.4_{-2.6}^{+1.1}$  & $161.8_{-13.5}^{+12.1}$\\\\
10-20  & 100-71  & $72.1_{-1.9}^{+0.6}$  & $107.5_{-8.6}^{+6.3}$\\\\
20-30  & 70-49   & $51.8_{-1.2}^{+0.5}$  & $68.4_{-4.7}^{+3.6}$\\\\
30-40  & 48-33   & $36.2_{-0.8}^{+0.4}$  & $42.3_{-2.6}^{+1.9}$\\\\
40-50  & 32-22   & $24.9_{-0.6}^{+0.4}$  & $25.9_{-1.5}^{+1.0}$\\\\
50-60  & 21-14   & $16.3_{-0.3}^{+0.4}$  & $15.1_{-0.6}^{+0.6}$\\\\
\end{tabular}
\end{ruledtabular}
\end{table}

\subsection{Event and Centrality Selection}
\label{eventcut}

   For this analysis we selected events with a collision vertex within $\pm$30 cm measured 
along the beam axis from the center of the TPC. 
This event selection is applied to all the data sets discussed here.

   The events are further binned according to collision centrality
which is determined by the measured charged hadron multiplicity within
the pseudo-rapidity range $|{\eta}|$ $<$ 0.5.
In Table~\ref{tab:table1} we list the centrality bins for 
Au+Au at $\sqrt{s_{\rm{NN}}}$ = 62.4 GeV along with the multiplicity bin
definitions, average number of participating nucleons and average number of binary 
nucleon-nucleon collisions~\cite{Adams:2005cy,Miller:2007ri}.
For the present analysis we chose six centrality bins corresponding
to 0-5$\%$, 5-10$\%$, 10-20$\%$, 20-30$\%$, 30-50$\%$, 50-80$\%$ of the total 
inelastic nucleus-nucleus hadronic cross-section. A dataset of 2 million
minimum-bias trigger events which passed the event cuts 
is used in the analysis.

Tables~\ref{tab:table2} and \ref{tab:table3} list
the six centrality bins for Cu+Cu at $\sqrt{s_{\rm{NN}}}$ = 200 GeV
and 62.4 GeV corresponding to 0-10$\%$, 10-20$\%$, 20-30$\%$, 30-40$\%$, 40-50$\%$, 50-60$\%$ 
of the total hadronic cross-section. The number of events used is
15 million and 24 million for 62.4~GeV and 200~GeV Cu+Cu datasets, respectively, after 
the event cuts.

\subsection{Particle Selection}
\label{trackcut}

   We selected particle tracks in the rapidity region $|y|$ $<$ 0.5. Particle identification 
was performed by correlating the specific ionization of particles in the TPC gas with their measured 
momenta. For this analysis pions are selected by requiring the specific ionization to be within 
2 standard deviations from their theoretical Bichsel value~\cite{Bichsel:2006cs,abelevstar:2008ez}. In order to remove the kaons and 
protons which could satisfy this condition, particles are also required to be more than 2 
standard deviations from the Bethe-Bloch value for kaons and protons. 
Charged particle tracks reconstructed and used for this analysis are accepted 
if they have space points on at least 15 pad rows in TPC. Tracks with fewer 
space points may be broken track fragments.
These cuts are similar to those in our previous analysis of Au+Au collisions at 
$\sqrt{s_{\rm{NN}}}$ = 200 GeV~\cite{Adams:2004yc} since the detector setup was identical.

\subsection{Pair Cuts}
\label{paircut}

Two types of particle track reconstruction errors directly affect measured
particle pair densities at the small relative momentum values studied here.
Track splitting, in which one particle trajectory is reconstructed as two or
more ``particles,'' increases the apparent number of pairs at low relative $q$. 
To address this problem we developed a split track filter algorithm, described
in our previous analysis of Au+Au collisions at $\sqrt{s_{\rm{NN}}}$ =
200 GeV~\cite{Adams:2004yc}, where values of the splitting level parameter
from $-$0.5 to 0.6~\cite{Adams:2004yc} ensured valid tracks. The inefficiencies arising due to track merging, 
in which two or more particle trajectories are reconstructed as one track,
was completely eliminated by requiring that the
fraction of merged hits (overlapping space-charge depositions in the TPC gas)
be less than 10$\%$ for every track pair used in the correlation function.

In the present analysis, we used the same cuts to remove splitting and merging as were
used for Au+Au collisions at $\sqrt{s_{\rm{NN}}}$ = 200 GeV~\cite{Adams:2004yc}.
The track pairs are required to have an average transverse momentum 
($k_{\rm T}$ $=$ ($|\vec{p}_{\rm 1T}$ $+$ $\vec{p}_{\rm 2T}|$)$/$2) in one of 4 bins corresponding to 
[150,250] MeV/c, [250,350] MeV/c, [350,450] MeV/c and [450,600] MeV/c. The results are presented and discussed 
as a function of $k_{\rm T}$ and $m_{\rm T}$ (= $\sqrt{k_{\rm T}^{2}+m_{\rm \pi}^{2}}$) in each of those bins.

\section{Analysis Method}
\label{analysis}
\subsection{Correlation function}

\label{corrfunc}       

The numerator and denominator of the two particle correlation function
in Eq.(\ref{eq:one}) are constructed by filling histograms corresponding
to particle pairs from the same event and from mixed events, respectively.
The background pairs are constructed 
from mixed events~\cite{Kopylov:1972qw} where by pairing each particle in a given event is mixed with all particles 
from other events within a subset of ten similar events. The events for mixing are selected
within the given centrality bin such that their respective primary vertex
$\it{z}$ positions are all within 10 cm of one another. 

\subsection{Bertsch-Pratt Parametrizations and Coulomb interactions}
\label{BPparam}

We decompose the relative momentum $\overrightarrow{q}$ according to the 
Bertsch-Pratt (or ``out-side-long'') convention~\cite{Podgoretsky,Grassberger,Bertsch:1988db,Pratt:1986cc,Chapman:1994yv}. 
The relative momentum $\overrightarrow{q}$ is decomposed into the variables $\it{q_{\rm long}}$
 along the beam direction, $\it{q_{\rm out}}$ parallel to the transverse momentum of the pair 
$\vec{k}_{\rm T}$ $=$ ($\vec{p}_{\rm 1T}$ $+$ $\vec{p}_{\rm 2T}$)/2, 
and $\it{q_{\rm side}}$ perpendicular to $\it{q_{\rm long}}$ and $\it{q_{\rm out}}$.

In addition to the correlation arising from the quantum statistics of two
identical (boson) particles, correlations can also arise from two-particle final state interactions even  
for non-identical particles~\cite{lednicky,Gyulassy:1979yi,Boal:1990yh}. For identical pions the effects 
of strong interactions are negligible, but the long range Coulomb repulsion causes a 
suppression of the measured correlation function 
at small $\overrightarrow{q}$.

In this paper we follow the procedure used in our previous analysis 
of Au+Au collisions at $\sqrt{s_{\rm{NN}}}$ = 200 GeV~\cite{Adams:2004yc}. For an azimuthally
integrated analysis at mid-rapidity in the longitudinal co-moving system (LCMS) the correlation 
function in Eq.~(\ref{eq:one}) can be decomposed as~\cite{Lisa:2005dd,Sinyukov:1998fc}: 
\begin{equation}
\label{eq:two}
C(q_{\rm out},q_{\rm side},q_{\rm long}) = ( 1 - \lambda )  + ~~~~~~~~~~~~~~~~~~~~~~~~~~~~~~~~~~~~~~~~~~~~~~~~~~~~~~~~  \nonumber
\end{equation}
\begin{equation}
\lambda K_{\rm coul}(q_{\rm inv})(1 + e^{-q^2_{\rm out}R^2_{\rm out} - q^2_{\rm side}R^2_{\rm side} - q^2_{\rm long}R^2_{\rm long}}),
\end{equation}
where $K_{\rm coul}$ is, to a good approximation, the squared nonsymmetrized Coulomb wave function
integrated over a Gaussian source
(corresponding to the LCMS Gaussian radii $R_{\rm out}$, $R_{\rm side}$, $R_{\rm long}$).
Assuming perfect experimental particle identification and a purely chaotic (incoherent) source, lambda represents the fraction
of correlated pairs~\cite{Lisa:2008gf}.

We assumed a spherical Gaussian source of 5 fm for Au+Au collisions at $\sqrt{s_{\rm{NN}}}$ = 62.4 GeV and 
a 3 fm source for Cu+Cu collisions at $\sqrt{s_{\rm{NN}}}$ = 62.4 and 200 GeV.
The first term (1 - $\lambda$) in Eq.(\ref{eq:two}) accounts for those 
pairs which do not interact or interfere and the second term represents those pairs where both 
Bose-Einstein effects and Coulomb interactions are present~\cite{Adams:2004yc}. 

\subsection{Systematic Uncertainties}
\label{SysErrors}

We studied several sources of systematic errors
similar to a previously published STAR pion interferometry analysis 
for Au+Au collisions at $\sqrt{s_{\rm{NN}}}$ = 200 GeV~\cite{Adams:2004yc}. 
The following effects are considered: track merging, track splitting, 
source size assumed for the Coulomb correction, particle identification 
purity, and particle pair acceptance effects for unlike-sign charged pions.
The estimated systematic errors are less than 10$\%$ for
$R_{\rm out}$, $R_{\rm side}$, $R_{\rm long}$, $\lambda$ in all centrality and $k_{\rm T}$ bins for the present
datasets and are similar to those in~\cite{Adams:2004yc}. 
This similarity is expected since the detector setup was identical and similar particle and
pair selection cuts are used for Au+Au and Cu+Cu
collisions. Results shown in the figures for the present datasets include statistical
errors only.

\section{Results and Discussion}
\label{results}

\subsection{Au+Au collisions at $\sqrt{s_{\rm{NN}}}$ = 62.4 GeV}
\label{auau62}

\begin{figure}
\includegraphics[width=20pc]{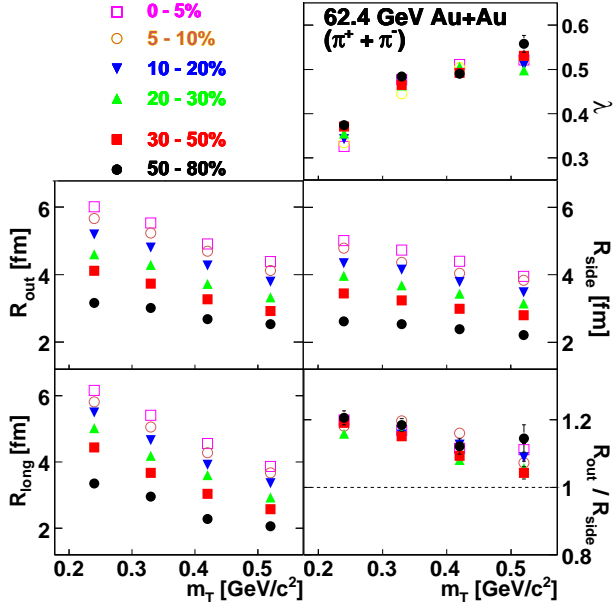}% Here is how to import EPS art
\caption{\label{fig:Figure1}(Color Online) The femtoscopic parameters vs. $m_{\rm T}$ for 6 different 
centralities for Au+Au collisions at $\sqrt{s_{\rm{NN}}}$ = 62.4 GeV. 
Only statistical errors are shown. The estimated systematic errors are less than 10$\%$ for 
$R_{\rm out}$, $R_{\rm side}$, $R_{\rm long}$, $\lambda$ in all centrality and $k_{\rm T}$ bins.}
\end{figure}

\begin{figure}
\includegraphics[width=20pc]{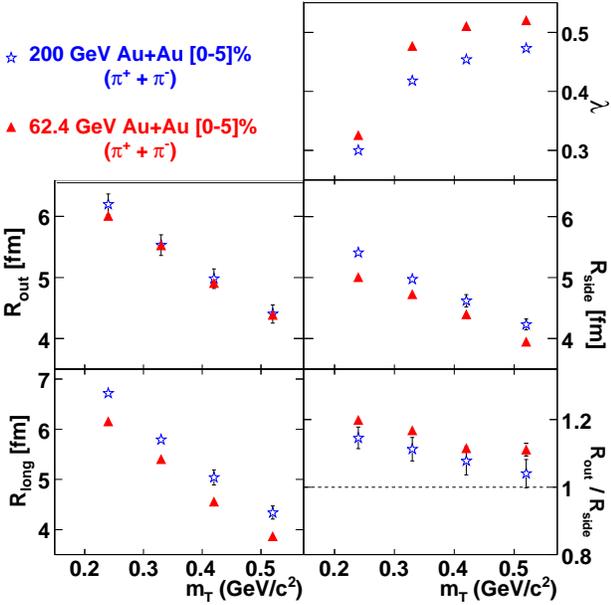}% Here is how to import EPS art
\caption{\label{fig:Figure2}(Color Online) The comparison of femtoscopic measurements of Au+Au 
collisions at $\sqrt{s_{\rm{NN}}}$ = 200 GeV and 62.4 GeV for 0-5$\%$ most 
central events. Only statistical errors are shown for 
Au+Au collisions at $\sqrt{s_{\rm{NN}}}$ = 62.4 GeV. The estimated systematic errors for 
Au+Au collisions at $\sqrt{s_{\rm{NN}}}$ = 62.4 GeV are less than 10$\%$ for $R_{\rm out}$, $R_{\rm side}$, $R_{\rm long}$, $\lambda$ in 0-5$\%$ most 
central events and all $k_{\rm T}$ bins. 
The 200 GeV results are from~\cite{Adams:2004yc}.}
\end{figure}

\begin{figure}
\includegraphics[width=20pc]{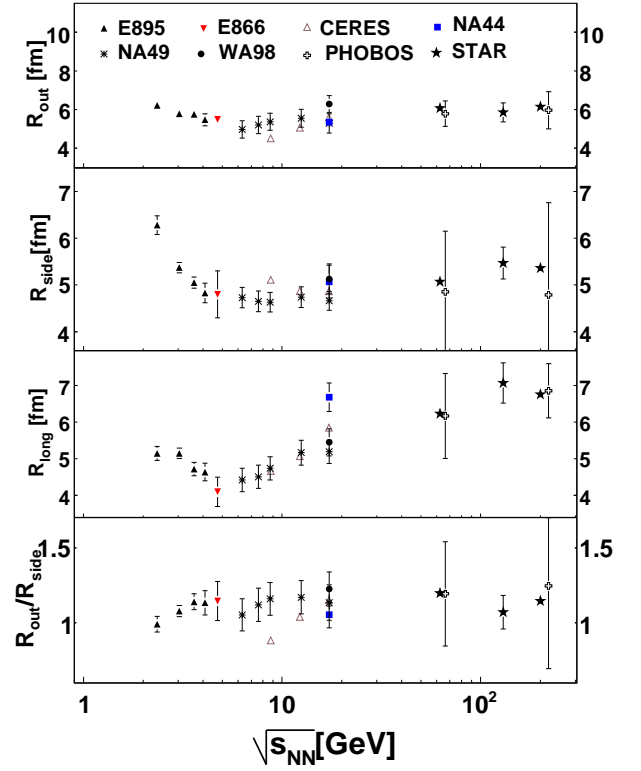}% Here is how to import EPS art
\caption{\label{fig:Figure3}(Color Online) The energy dependence of femtoscopic parameters for 
AGS, SPS and RHIC from Refs.~\cite{Adler:2001zd,Adams:2004yc,Back:2004ug,Ahle:2002mi,Adamova:2002wi,Lisa:2000hw,
Bearden:1998aq,Alt:2007uj,Soltz:1999iy,Aggarwal:2002tm}. 
Energy dependences of pion femtoscopic parameters for central Au+Au, Pb+Pb and Pb+Au 
collisions are shown for mid-rapidity and $\langle k_T \rangle \sim$ 0.2-0.3 GeV/c. 
Error bars on NA44, NA49, CERES, PHOBOS and STAR results at 
$\sqrt{s_{\rm{NN}}}$ = 130 and 200 GeV include systematic 
uncertainties; error bars on other results are statistical only.
Only statistical errors are shown for 
Au+Au collisions at $\sqrt{s_{\rm{NN}}}$ = 62.4 GeV; the estimated systematic errors 
are less than 10$\%$ for $R_{\rm out}$, $R_{\rm side}$, $R_{\rm long}$.
The PHOBOS results from ~\cite{Back:2004ug} 
for $\sqrt{s_{\rm{NN}}}$ = 62.4 and 200 GeV are slightly 
shifted horizontally for visual clarity.}
\end{figure}

The correlation function in  Eq.(\ref{eq:two}) is fitted to the 3D correlation data for
Au+Au collisions at $\sqrt{s_{\rm{NN}}}$ = 62.4 GeV for each centrality
and $m_{\rm T}$ bins as defined above. The analysis is performed separately for
$\pi^{+}\pi^{+}$ and $\pi^{-}\pi^{-}$ pairs. The final histograms 
for the like-sign pairs do not show appreciable differences and may therefore be
summed in order to increase statistics. 
Figure~\ref{fig:Figure1} presents the results for 
$R_{\rm out}$, $R_{\rm side}$, $R_{\rm long}$,
$\lambda$ and the ratio, $R_{\rm out}$$/$$R_{\rm side}$. 
The three femtoscopic radii increase with increasing centrality as expected, whereas the values of
$\lambda$ and the $R_{\rm out}$$/$$R_{\rm side}$ ratio 
exhibit no clear centrality dependences. 

We observe that for all centralities 
the three femtoscopic radii decrease 
with increasing $m_{\rm T}$
whereas the $\lambda$ parameter increases with $m_{\rm T}$. Such behavior is 
consistent with our previous measurements at 
$\sqrt{s_{\rm{NN}}}$ = 200 GeV~\cite{Adams:2004yc}.
The increase of parameter $\lambda$ with  $m_{T}$ is due to the
decreasing contribution of pions produced from long-lived resonance decays at higher transverse
momenta. For comparison, in Fig.~\ref{fig:Figure2} we show the results for Au+Au collisions at 
$\sqrt{s_{\rm{NN}}}$ = 62.4 GeV and 200 GeV for the most central collisions.
We observe that the $R_{\rm out}$ values are similar for both cases, but
there are differences between the values of $R_{\rm side}$ and $R_{\rm long}$. 
The $R_{\rm out}$$/$$R_{\rm side}$ ratio decreases with increasing $m_{\rm T}$, but the values
are higher for $\sqrt{s_{\rm{NN}}}$  = 62.4 GeV than 
for $\sqrt{s_{\rm{NN}}}$ = 200 GeV.

The observed dependences of the three femtoscopic radii are
qualitatively consistent with models with collective 
flow~\cite{Kolb:2003dz,Retiere:2003kf,Hirano:2002hv}. 
Collective expansion results in position-momentum correlations in 
both transverse and longitudinal directions. 
In an expanding source the correlation between the space-time points where the pions 
are emitted and their energy-momentum produce a characteristic dependence of 
femtoscopic radii on $m_{\rm T}$~\cite{Pratt:1986cc,Heinz:1996bs,
Wiedemann:1997cr,Pratt:1984su,Tomasik:1999ct,Wiedemann:1995au,
Schlei:1996mc,Schlei:1992jj,Lisa:2005dd}.
The decrease in the ``out'' and ``side'' components can be described by models including transverse flow
~\cite{Heinz:1996bs,Wiedemann:1997cr,Tomasik:1999ct,Adams:2004yc}, 
and the decrease in the ``long'' component by those with longitudinal
flow~\cite{Makhlin:1987gm,Wiedemann:1997cr,
Wiedemann:1995au,Adams:2004yc}.

\subsection{Energy dependence of femtoscopic radii} 
\label{centdep}

In Fig.~\ref{fig:Figure3} we present the energy dependences of
the three femtoscopic radii and the ratio
$R_{\rm out}$$/$$R_{\rm side}$ for the available data
from AGS, SPS and RHIC. The results are compiled for Au+Au, Pb+Pb
and Pb+Au collisions at mid-rapidity and for  $\langle k_{\rm T} \rangle \sim$ 0.2-0.3 GeV/c.
The present measurements for Au+Au collisions at $\sqrt{s_{\rm{NN}}}$
= 62.4 GeV are also included.
The quality of the present STAR data with respect to statistical and
systematic errors is significantly better than that reported by PHOBOS~\cite{Back:2004ug} at the same energy.
PHENIX results are not included because they were reported for broader centrality bins. 
WA97 results are also omitted because they were measured at higher transverse momenta.

Comparative studies are a necessary part of searches for nontrivial structures in the excitation function which might arise from
a possible phase transition~\cite{Rischke:1996em}.
The radius parameter $R_{\rm side}$ has the most direct correlation with the source
geometry whereas $R_{\rm out}$ encodes both geometry and time scale information. Experimental results
show that $R_{\rm side}$ decreases at AGS energies and then displays a modest rise with collision energy
from SPS to RHIC. $R_{\rm long}$ increases with collision energy after an initial decrease  
at the lower AGS energies. For $R_{\rm out}$ the changes are very small. 

Hydrodynamic model calculations ~\cite{Rischke:1996em,Rischke:1996nq}
predict an enhancement in the ratio of $R_{\rm out}/R_{\rm side}$
with increasing beam energy. The experimental results show no
such behavior. The measured ratios are better reproduced by the
AMPT (A Multi-Phase Transport) model~\cite{Lin:2002gc},
however the individual predicted radii have a steeper decrease compared to the experimental
data \cite{Lisa:2005dd}. An alternative model using a
relativistic quantum mechanical treatment of opacity and the refractive 
index is capable of reproducing the observed results~\cite{Cramer:2004ih}, but strongly 
depends on the assumed initial conditions and neglects the time dependence of the
corresponding optical potential. Hydrodynamic calculations~\cite{Teaney:2003kp} 
including viscosity offer another possible explanation for the above deviation between the data and model
calculations as recently shown in~\cite{Romatschke:2007jx}.
According to recent hydrodynamic calculations, the femtoscopic
radii can be described either by using the initial Gaussian
density profile~\cite{Florkowski:2008xy} or by including the combination of
several effects including: pre-thermal acceleration, a stiffer equation
of state, and additional viscous corrections~\cite{Pratt:2008qv}.
Other recent studies with a granular source model~\cite{Zhang:2007wg} 
also obtain a better description of the experimental measurements of pion femtoscopic radii. 

\subsection{Cu+Cu collisions at $\sqrt{s_{\rm{NN}}}$ = 62.4 and 200 GeV} 
\label{CuCu_compare}

\begin{figure}
\includegraphics[width=20pc]{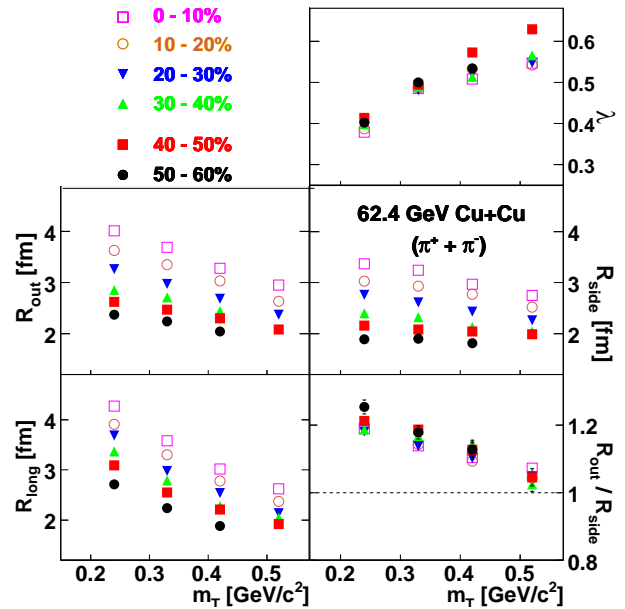}% Here is how to import EPS art
\caption{\label{fig:Figure4}(Color Online) Femtoscopic parameters vs. $m_{T}$ for six centralities 
for Cu+Cu collisions at $\sqrt{s_{\rm{NN}}}$ = 62.4 GeV. Only statistical errors are shown. The estimated systematic errors 
are less than 10$\%$ for $R_{\rm out}$, $R_{\rm side}$, $R_{\rm long}$, $\lambda$ in all centrality and $k_{\rm T}$ bins.}
\end{figure}

\begin{figure}
\includegraphics[width=20pc]{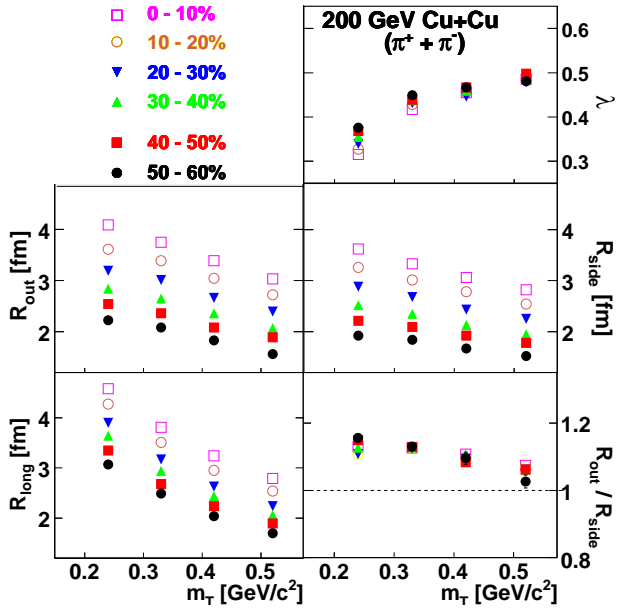}% Here is how to import EPS art
\caption{\label{fig:Figure5}(Color Online) Femtoscopic parameters vs. $m_{T}$ for six centralities for 
Cu+Cu collisions at $\sqrt{s_{\rm{NN}}}$ = 200 GeV. Only statistical errors are shown. The estimated systematic errors 
are less than 10$\%$ for $R_{\rm out}$, $R_{\rm side}$, $R_{\rm long}$, $\lambda$ in all centrality and $k_{\rm T}$ bins.}
\end{figure}

The correlation functions are similarly constructed for Cu+Cu collisions
at $\sqrt{s_{\rm{NN}}}$ = 62.4 GeV and 200 GeV. The extracted femtoscopic radii, 
$R_{\rm out}$, $R_{\rm side}$ and $R_{\rm long}$, along with the $\lambda$ parameter and the ratio $R_{\rm out}$$/$$R_{\rm side}$
are presented in Figs.~\ref{fig:Figure4} and \ref{fig:Figure5}
for the 62.4 and 200 GeV data, respectively.
The results are presented for six different centralities and four $m_{\rm T}$ bins. The highest $k_{\rm T}$ bin [450 - 600] MeV/c of the 
most peripheral centrality (50 - 60 $\%$) in Cu+Cu collisions
at $\sqrt{s_{\rm{NN}}}$ = 62.4 GeV is omitted due to inadequate statistics for decomposition with the Bertsch-Pratt 
parametrization.
For both collision energies the three femtoscopic radii increase with increasing centrality
whereas the $\lambda$ parameter shows no centrality dependence.
The $m_{\rm T}$ dependences of the femtoscopic radii are similar to that for Au+Au collisions.
The $R_{\rm out}/R_{\rm side}$ ratios exhibit no clear centrality 
dependences for either energy.

\subsection{Comparison of femtoscopic radii for Cu+Cu and Au+Au collisions} 
\label{Au_Cu_compare}

\begin{figure}
\includegraphics[width=20pc]{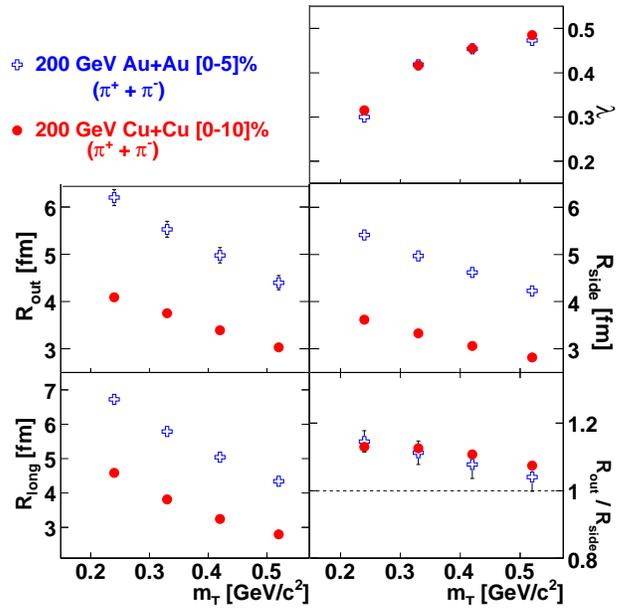}% Here is how to import EPS art
\caption{\label{fig:Figure6}(Color Online) The comparison of system size dependence in femtoscopic measurements of STAR Au+Au and Cu+Cu 
collisions at $\sqrt{s_{\rm{NN}}}$ = 200 GeV. Only statistical errors are shown for 
Cu+Cu collisions at $\sqrt{s_{\rm{NN}}}$ = 200 GeV. The estimated systematic errors for 
Cu+Cu collisions at $\sqrt{s_{\rm{NN}}}$ = 200 GeV are less than 10$\%$ for $R_{\rm out}$, $R_{\rm side}$, $R_{\rm long}$, $\lambda$ in 0-10$\%$ most 
central events and $k_{\rm T}$ bins. 
The Au+Au results are from~\cite{Adams:2004yc}.}
\end{figure}

In Fig.~\ref{fig:Figure6} the femtoscopic source parameters
$\lambda$, $R_{\rm out}$,  $R_{\rm side}$,  $R_{\rm long}$ and  the ratio $R_{\rm out}$$/$$R_{\rm side}$
for central (0-5$\%$) Au+Au collisions at $\sqrt{s_{\rm{NN}}}$ = 200 GeV~\cite{Adams:2004yc} 
are compared with central (0-10$\%$) Cu+Cu collisions at same beam energy. 
As expected, the femtoscopic radii for Cu+Cu collisions are smaller 
than for Au+Au collisions at the same beam energy. It is interesting
that the values of the ratio $R_{\rm out}$$/$$R_{\rm side}$ for the two systems are similar.

\begin{figure}
\includegraphics[width=20pc]{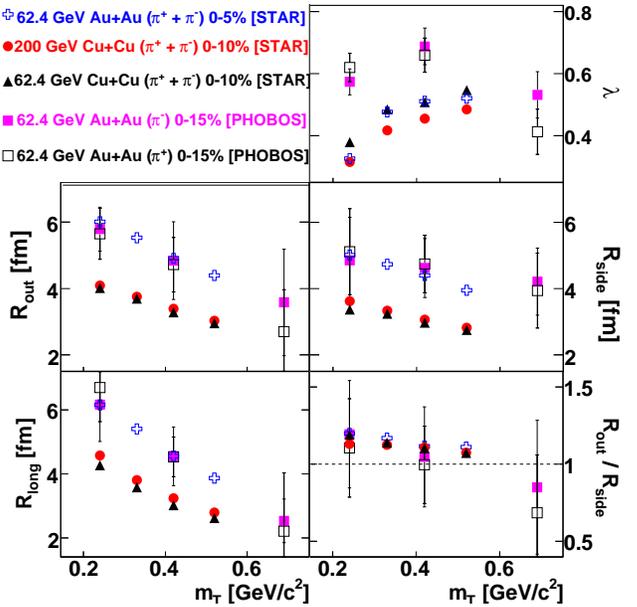}% Here is how to import EPS art
\caption{\label{fig:Figure7}(Color Online) The comparison of femtoscopic measurements of STAR Cu+Cu collisions 
at $\sqrt{s_{\rm{NN}}}$ = 200 and 62.4 GeV and Au+Au collisions at $\sqrt{s_{\rm{NN}}}$ = 62.4 GeV. Only statistical errors are shown for 
STAR results. The estimated systematic errors for STAR results are less than 10$\%$ for $R_{\rm out}$, $R_{\rm side}$, $R_{\rm long}$, $\lambda$ in all 
centrality and $k_{\rm T}$ bins. The PHOBOS results~\cite{Back:2004ug} for positive and negative pions in 
Au+Au collisions at $\sqrt{s_{\rm{NN}}}$ = 62.4 GeV are compared with STAR results.}
\end{figure}

In Fig.~\ref{fig:Figure7} we extend the comparison 
of femtoscopic source parameters to include central (0-5$\%$) Au+Au collisions at 
$\sqrt{s_{\rm{NN}}}$ = 62.4 GeV, central (0-10$\%$) Cu+Cu collisions 
at $\sqrt{s_{\rm{NN}}}$ = 62.4 and 200 GeV, and central (0-15$\%$) $\pi^+ \pi^+$
and $\pi^- \pi^-$ correlations from Au+Au collisions at 62.4~GeV from the
PHOBOS experiment~\cite{Back:2004ug}.
The femtoscopic radii for Cu+Cu collisions at $\sqrt{s_{\rm{NN}}}$ = 62.4 GeV are smaller 
than those for Au+Au collisions at the same beam energy. 
The femtoscopic radii for Cu+Cu central collisions are similar for both energies. 
The variation of the $R_{\rm out}/R_{\rm side}$ ratio with $m_{\rm T}$ is similar for the
Au+Au and Cu+Cu collision data.

\begin{figure}
\includegraphics[width=20pc]{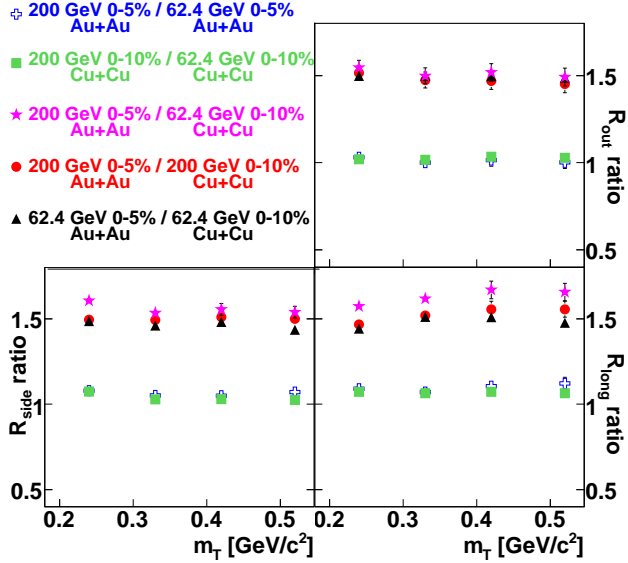}% Here is how to import EPS art
\caption{\label{fig:Figure8}(Color Online) Ratios of femtoscopic radii at top centralities for Au+Au and Cu+Cu 
collisions at $\sqrt{s_{\rm{NN}}}$ = 200 and 62.4 GeV vs. $m_{T}$. Only statistical errors are shown for 
Au+Au collisions at $\sqrt{s_{\rm{NN}}}$ = 62.4 GeV and Cu+Cu collisions at $\sqrt{s_{\rm{NN}}}$ = 62.4 
and 200 GeV. The estimated systematic errors for Au+Au collisions at $\sqrt{s_{\rm{NN}}}$ = 62.4 GeV and Cu+Cu 
collisions at $\sqrt{s_{\rm{NN}}}$ = 62.4 and 200 GeV are less than 10$\%$ for $R_{\rm out}$, $R_{\rm side}$, $R_{\rm long}$ in all centrality and $k_{\rm T}$ bins. 
The 200 GeV results are from~\cite{Adams:2004yc}.}
\end{figure}

In Fig.~\ref{fig:Figure8} we present the $m_{\rm T}$ dependences of the ratios of femtoscopic radii for the most-central Au+Au and Cu+Cu collisions at 
$\sqrt{s_{\rm{NN}}}$ = 200 and 62.4~GeV.
Ratios for the same colliding ion systems are close to unity whereas ratios of radii for Au+Au to Cu+Cu collisions are $\sim$1.5.
Although the individual radii decrease significantly with increasing $m_{\rm T}$
the ratios in Fig.~\ref{fig:Figure8} show that the femtoscopic radii for 
Au+Au and Cu+Cu collisions at 62.4 and 200~GeV share a common $m_{\rm T}$ dependence.
This result can be understood in terms of models~\cite{Cramer:2004ih,Miller:2005ji} 
which use participant scaling to predict the femtoscopic radii in Cu+Cu collisions
from the measured radii for Au+Au collisions at $\sqrt{s_{\rm{NN}}}$ = 200 GeV, 
assuming the radii are proportional to
A$^{1/3}$, where A is the atomic mass number of the colliding nuclei.

\subsection{Volume estimates and multiplicity scaling}

Estimates of the pion freeze-out volume $V_{f}$ in terms of the femtoscopic radii
are provided by the following expressions:
\begin{subequations}
\begin{equation}
\label{eq:five1}
V_{f} \propto R_{\rm side}^2R_{\rm long}
\end{equation}
\begin{eqnarray}
\label{eq:five2}
~~~~~~~~~~~V_{f} \propto R_{\rm out}R_{\rm side}R_{\rm long}.
\end{eqnarray}
\end{subequations}
However, the correlation lengths (femtoscopic radii) decrease with increasing $m_{\rm T}$ corresponding to an $m_{\rm T}$ 
dependent region of homogeneity which, in expanding source models, is smaller than the true collision volume at freeze-out.
The volume estimates 
[Eqs.~(\ref{eq:five1}) and (\ref{eq:five2})] are obtained from the lowest $m_{\rm T}$ bin, corresponding to the 
$k_{\rm T}$ region from 150 to 250 MeV/c as discussed in Sec.~(\ref{paircut}).

\begin{figure}
\includegraphics[width=20pc]{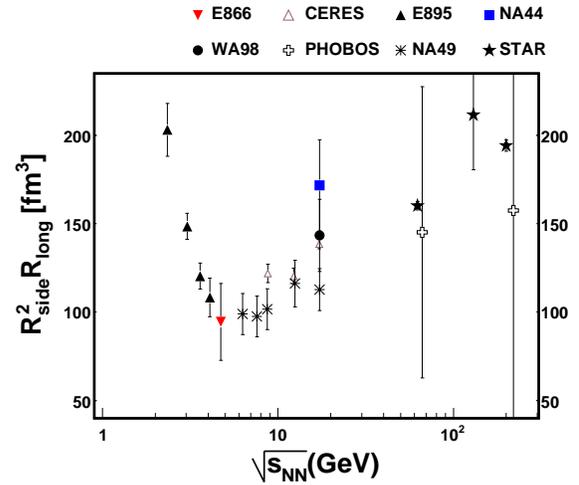}% Here is how to import EPS art
\caption{\label{fig:Figure9a}(Color Online) The energy dependence of the
pion freeze-out volume for heavy ion collision data from the AGS, SPS and RHIC estimated using 
Eq.~(\ref{eq:five1}). The references are given in the caption of Fig.~\ref{fig:Figure3}. Only statistical errors are shown for 
Au+Au collisions at $\sqrt{s_{\rm{NN}}}$ = 62.4 GeV. The estimated systematic errors for 
Au+Au collisions at $\sqrt{s_{\rm{NN}}}$ = 62.4 GeV are less than 10$\%$ for $R_{\rm out}$, $R_{\rm side}$, $R_{\rm long}$ in all centrality and $k_{\rm T}$ bins.}
\end{figure}

\begin{figure}
\includegraphics[width=20pc]{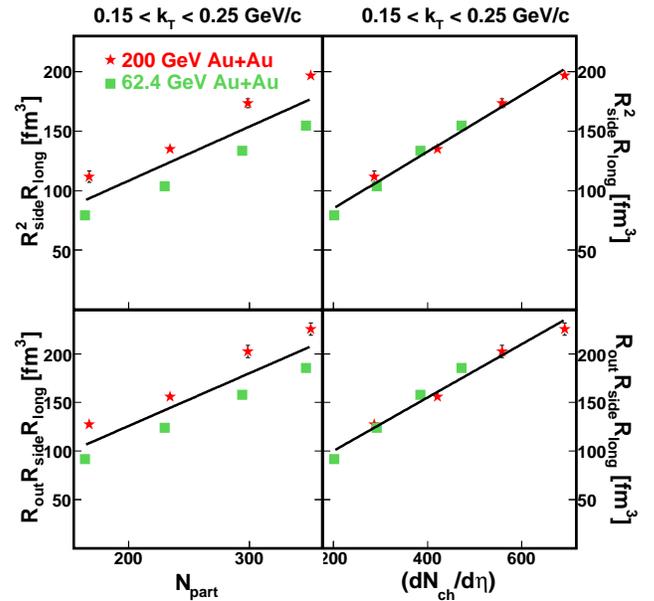}% Here is how to import EPS art
\caption{\label{fig:Figure9}(Color Online) Pion freeze-out volume estimates as a function of number 
of participants and charged particle multiplicity density for Au+Au at 
$\sqrt{s_{\rm{NN}}}$ = 200 and 62.4 GeV. Only statistical errors are shown for 
Au+Au collisions at $\sqrt{s_{\rm{NN}}}$ = 62.4 GeV. The estimated systematic errors for 
Au+Au collisions at $\sqrt{s_{\rm{NN}}}$ = 62.4 GeV are less than 10$\%$ for $R_{\rm out}$, $R_{\rm side}$, $R_{\rm long}$ in all centrality and $k_{\rm T}$ bins. 
The 200 GeV Au+Au collision results are from~\cite{Adams:2004yc}. The lines in each panel represent linear fits to the data.}
\end{figure}

\begin{figure}
\includegraphics[width=20pc]{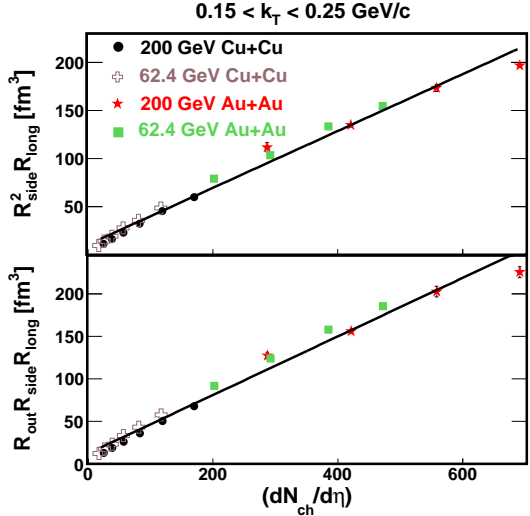}% Here is how to import EPS art
\caption{\label{fig:Figure10}(Color Online) Pion freeze-out volume estimates 
as a function of charged particle multiplicity density for Au+Au and Cu+Cu collisions. Only statistical errors are shown for 
Au+Au collisions at $\sqrt{s_{\rm{NN}}}$ = 62.4 GeV and Cu+Cu collisions at $\sqrt{s_{\rm{NN}}}$ = 62.4 
and 200 GeV. The estimated systematic errors for Au+Au collisions at $\sqrt{s_{\rm{NN}}}$ = 62.4 GeV and Cu+Cu collisions at $\sqrt{s_{\rm{NN}}}$ = 62.4 
and 200 GeV are less than 10$\%$ for $R_{\rm out}$, $R_{\rm side}$, $R_{\rm long}$ in all centrality and $k_{\rm T}$ bins. 
The 200 GeV Au+Au collision results are from~\cite{Adams:2004yc}. The lines in each panel represent linear fits to the data.}
\end{figure}

\begin{figure}
\begin{center}
\includegraphics[height=20pc]{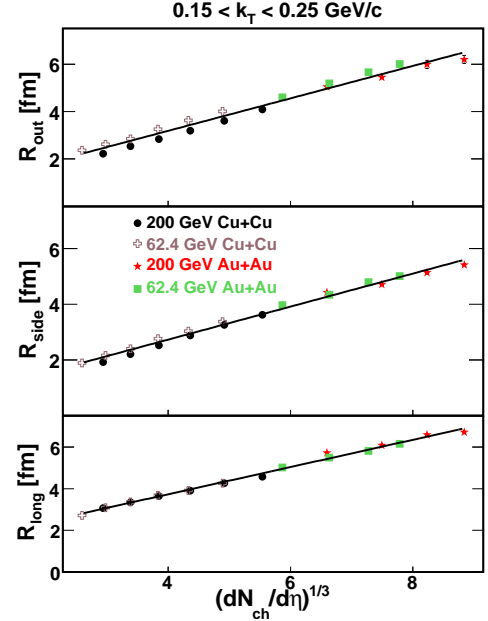}% Here is how to import EPS art
\end{center}
\caption{\label{fig:Figure11}(Color Online) The pion source radii dependences on charged 
particle multiplicity density for Au+Au and Cu+Cu collisions. Only statistical errors are shown for 
Au+Au collisions at $\sqrt{s_{\rm{NN}}}$ = 62.4 GeV and Cu+Cu collisions at $\sqrt{s_{\rm{NN}}}$ = 62.4 
and 200 GeV. The estimated systematic errors for Au+Au collisions at $\sqrt{s_{\rm{NN}}}$ = 62.4 GeV and Cu+Cu collisions at $\sqrt{s_{\rm{NN}}}$ = 62.4 
and 200 GeV are less than 10$\%$ for $R_{\rm out}$, $R_{\rm side}$, $R_{\rm long}$ in all centrality and $k_{\rm T}$ bins.  
The 200 GeV Au+Au collision results are from~\cite{Adams:2004yc}. 
The lines represent linear fits to the data.}
\end{figure}

The $V_{f}$ measurements using Eq.~(\ref{eq:five1}) as a function of $\sqrt{s_{\rm NN}}$ are
presented in Fig.~\ref{fig:Figure9a} for Au+Au, Pb+Pb and Pb+Au collisions at mid-rapidity 
and for the lowest $k_{\rm T}$ bin defined above.
The results show two distinct domains: First, at the AGS where the volume measure
decreases, and second, in the SPS and RHIC energy regimes where a monotonic increase is observed. 

A detailed description of this non-trivial behavior was suggested in ~\cite{Adamova:2002ff} 
based on the hypothesis of constant mean free path length of pions at freeze-out. 
The explanation provided in~\cite{Adamova:2002ff} defines the pion
mean free path length, $\lambda_{f}$, as:
\begin{equation}
\label{eq:six}
\lambda_{f} = \frac{1}{\rho_{f}\sigma} = \frac{V_{f}}{N\sigma},
\end{equation}
where $\rho_{f}$ is the freeze-out density and $\sigma$ is the total cross-section for 
pions to interact with the surrounding medium. The freeze-out density can be expressed as the 
number of particles $N$ in the estimated freeze-out volume $V_{f}$, divided by $V_{f}$, resulting 
in the second expression in Eq.~(\ref{eq:six}).  
The denominator, $N\sigma$, can be expanded as the sum of the pion-pion and pion-nucleon 
contributions. At AGS energies the pion-nucleon term dominates since the pion-nucleon 
cross-section is larger than the pion-pion cross-section. Also, the number of nucleons at these lower energies 
at mid-rapidity exceeds the number of pions. Hence, a decrease in the number of 
mid-rapidity nucleons leads to a decrease in the observed freeze-out volume ($V_{f}$)  
as a function of $\sqrt{s_{\rm{NN}}}$. At SPS and RHIC energies
the pion-pion term dominates the denominator in Eq.~(\ref{eq:six}) due to copious pion production leading to an 
increase in the observed $V_{f}$. 

Based on this interpretation we expect the 
volume estimates in the pion dominated RHIC regime to show a linear dependence on charged particle multiplicity.
In Fig.~\ref{fig:Figure9} freeze-out volume 
estimates (using Eqs.~(\ref{eq:five1}) and (\ref{eq:five2})) are shown as a function of 
the number of participants (left panels) and charged particle multiplicity (right panels) for Au+Au collisions
at $\sqrt{s_{\rm{NN}}}$ = 62.4 and 200 GeV. The predicted linear increase with
charged particle multiplicity is observed.
Estimated freeze-out volumes for Au+Au collisions at the same centralities
increase with collision energy indicating that $N_{\rm part}$ is not a suitable scaling variable in this case.
On the other hand, charged particle multiplicity provides better scaling properties.

Additional estimates of freeze-out volume dependences on 
charged particle multiplicity are presented in Fig.~\ref{fig:Figure10} for both
the Au+Au and Cu+Cu results at $\sqrt{s_{\rm{NN}}}$ = 62.4 and 200 GeV.
Both freeze-out volume estimates for the four collision systems 
show an approximate, common linear dependence on charged particle multiplicity.
The linear dependences of femtoscopic radii on $(dN_{\rm ch}/d\eta)^{1/3}$  
for Au+Au and Cu+Cu collisions at $\sqrt{s_{\rm{NN}}}$ = 62.4 and 200 GeV 
are shown in Fig.~\ref{fig:Figure11}. The above common, linear dependences~\cite{Lisa:2005dd} are consistent with the assumption 
of a universal pion mean-free-path length at freeze-out~\cite{Adamova:2002ff}.

\section{Summary and Conclusions}
\label{summary}

We have presented systematic measurements of pion femtoscopy for Au+Au collisions
at $\sqrt{s_{\rm{NN}}}$ = 62.4 GeV and Cu+Cu collisions at $\sqrt{s_{\rm{NN}}}$
= 62.4 and 200 GeV, and compared these new results with our previous analysis of Au+Au collisions at 
$\sqrt{s_{\rm{NN}}}$ = 200 GeV~\cite{Adams:2004yc}.
For all the systems considered the three femtoscopic radii ($R_{\rm out}$, $R_{\rm side}$ 
and $R_{\rm long}$) increase with centrality, whereas the values of the $\lambda$ parameter 
and ratio $R_{\rm out}$$/$$R_{\rm side}$ are approximately constant with centrality. 
The three femtoscopic radii decrease with increasing $m_{\rm T}$,
whereas the $\lambda$ parameter increases with $m_{\rm T}$. 
The increase of $\lambda$ with  $m_{\rm T}$ is attributed to
decreasing contamination from pions produced from long-lived resonance decays at higher transverse momentum.

The decrease of femtoscopic radii with increasing $m_{\rm T}$ can be described by models with collective, transverse and 
longitudinal expansion or flow. The ratios of femtoscopic radii at top centralities for
different colliding systems (Au+Au and Cu+Cu) at $\sqrt{s_{\rm{NN}}}$ = 62.4 and 200 GeV show that the
corresponding radii vary similarly with  $m_{\rm T}$.

The predicted rise of the ratio $R_{\rm out}/R_{\rm side}$ with collision energy due to a possible phase transition~\cite{Rischke:1996em} is not
observed for Au+Au and Cu+Cu collisions.
The compilation of freeze-out volume estimates $V_{f}$ as a function of collision energy $\sqrt{s_{NN}}$ (using Eq.~(\ref{eq:five1}) along 
with the datasets presented in Fig.~\ref{fig:Figure3}) shows two distinct domains: with increasing $\sqrt{s_{\rm{NN}}}$, $V_{f}$
decreases at the AGS, but steadily increases throughout the SPS and RHIC energy regimes.
At AGS energies the decreasing number of baryons at mid-rapidity leads to a decrease in the observed freeze-out volume ($V_{f}$)
as a function of $\sqrt{s_{\rm{NN}}}$. At higher beam energies from SPS to RHIC copious and increasing pion production
causes the freeze-out volume to rise. 

The dependences of the freeze-out volume estimate on number
of participants and charged particle multiplicity are compared.
Measurements for Au+Au collisions at the same centralities, but different
energies yield different freeze-out volumes demonstrating that $N_{\rm part}$ is not a suitable
scaling variable. The freeze-out volume estimates for all four collision systems presented here show a linear
dependence on final charged particle multiplicity which is consistent
with the hypothesis of a universal mean-free-path length at freeze-out.

For the systems studied here the multiplicity and $k_{\rm T}$ dependences of the
femtoscopic radii are consistent with previously established trends at RHIC
and at lower energies.  The radii scale with the final state collision multiplicity which, in
a static model, is consistent with an hypothesized universal mean-free-path length at
freeze-out. This and similar studies
establish the baseline systematics against which to
compare future femtoscopic studies at the LHC~\cite{Alessandro:2006yt}.

\vspace{0.1in}

We thank the RHIC Operations Group and RCF at BNL, and the
NERSC Center at LBNL and the resources provided by the
Open Science Grid consortium for their support. This work
was supported in part by the Offices of NP and HEP within
the U.S. DOE Office of Science, the U.S. NSF, the Sloan
Foundation, the DFG cluster of excellence `Origin and Structure
of the Universe,'
CNRS/IN2P3, RA, RPL, and EMN of France, STFC and EPSRC
of the United Kingdom, FAPESP of Brazil, the Russian
Ministry of Sci. and Tech., the NNSFC, CAS, MoST, and MoE
of China, IRP and GA of the Czech Republic, FOM of the
Netherlands, DAE, DST, and CSIR of the Government of India and the Korea Sci.~\& Eng.~Foundation. 
We wish to thank Polish State Committee for Scientific Research, grant:
N202 013 31/0489.

\bibliography{paperv3}% Produces the bibliography via BibTeX.
{}

\end{document}